\documentstyle[prl,aps]{revtex}

\def\blacksquare{\vrule height 4pt width 3pt depth2pt}

\def\be{\begin{equation}}
\def\ee{\end{equation}}
\def\hcal{{\cal H}}
\def\dcal{{\cal D}}
\def\tr{\mathop{\rm tr}}

\begin{document}
\draft

\title{Continuity of relative entropy of entanglement}

\author{Matthew J. Donald$^{1,}$\cite{poczta1} and Micha\l{}
Horodecki$^{2,}$\cite{poczta2}}

\address{$^1$ The Cavendish Laboratory, Madingley Road,
Cambridge CB3 0HE, Great Britain\\
$^2$ Institute of Theoretical Physics and Astrophysics,
University of Gda\'nsk, 80--952 Gda\'nsk, Poland}

\maketitle

\begin{abstract} We show that an entanglement measure
called relative entropy of entanglement satisfies a strong
continuity condition. If two states are close to each other
then so are their entanglements per particle pair in this
measure. It follows in particular,  that the measure is
appropriate for the description of entanglement
manipulations in the limit of an infinite number of pairs
of particles.
\end{abstract}


Entanglement is a crucial parameter in modern quantum
information theory
\cite{Ekert,dense,Bennett_tel,Shor,Bennett_pur,huge}. It is
therefore desirable to investigate properties of the
functions that quantify entanglement (entanglement
measures) \cite{huge,RP,Knight,VP98}. A property that has
recently appeared to be an important characteristic of
entanglement measures is continuity
\cite{Vidal_mon,miary}. This is especially relevant in the
description of manipulations of entanglement in the regime
of large numbers of identically prepared entangled pairs
(that is, in the case of stationary, memoryless sources) as
for example in the case of distillation of entanglement
\cite{Bennett_pur}. In general, one is interested in the
conversion of
$m$ pairs of particles, with each pair in state $\varrho$,
into $n$ pairs in another state $\varrho'$ by means of
local quantum operations and classical communication (LQCC)
\cite{huge}. Of course perfect transformation
 \[
\varrho^{\otimes m} \rightarrow \varrho'^{\otimes n}
\] is usually impossible. Thus one permits imperfections and
requires only asymptotically perfect transformations: the
state
$\varrho^{\otimes m}$ is transformed into some state
$\varrho'_n$, that for large $n$ approaches
$\varrho^{\otimes n}$. In this case one is interested in
entanglement measures that attribute approximately the same
entanglement per pair both to $\varrho'_n$ and
$\varrho^{\otimes n}$
\be D(\varrho'^{\otimes n},\varrho_n')\mathop{\rightarrow}
\limits^{n\rightarrow\infty} 0 \quad
\Rightarrow\quad
 {1\over n} |E(\varrho'^{\otimes n})-E(\varrho'_n)|
 \mathop{\rightarrow}
\limits^{n\rightarrow\infty} 0
 \label{slabsza}
\ee where $D$ is a chosen metric. We speak here about
entanglement {\it per pair} (or entanglement {\it density})
because, in the limit of infinite numbers of pairs, one
must use intensive quantities \cite{RP} (as in the
thermodynamics of lattice systems). It appears that the
above continuity, even if imposed only for  pure
$\varrho'$, puts severe constraints on entanglement
measures: all the additive measures satisfying that
condition must coincide on pure states
\cite{RP,Vidal_mon} and must be confined between
entanglement of distillation and entanglement of formation
\cite{huge} for mixed states
\cite{miary}.

In this paper, we consider the very important entanglement
measure: relative entropy of entanglement
\cite{Knight,VP98}.  For a state $\sigma$ acting on a
Hilbert space $\hcal_A\otimes \hcal_B$, this is given by
\[ E_r(\sigma) =\inf_{\varrho\in\dcal}S(\sigma|\varrho)
\] where $S(\sigma|\varrho)=\tr
\sigma\log\sigma-\tr\sigma\log\varrho$ and $\dcal$ is the
set of separable (disentangled) states.  This measure
was proved \cite{VP98,Rains_bound} to be a tight bound for
distillable entanglement, the central parameter of
entanglement based quantum communication \cite{huge} (for
the most straightforward proof, see \cite{miary}). We show
that it satisfies the very strong continuity
requirement constituted  by the Fannes-type inequality
(proved originally for the von Neumann entropy
\cite{Fannes})
\be |E_r(\varrho)-E_r(\sigma)|\leq B\log\dim \hcal
\kern0.4mm ||\varrho-\sigma|| + C \eta(||\varrho-\sigma||)
\label{Fannes1}
\ee where $\varrho$ and $\sigma$ act on the Hilbert space
$\hcal$,
$B$ and $C$ are constants, $\eta(s) = - s \log s$, and we
use the trace norm as a metric on states:
$D(\varrho,\sigma) = ||\varrho-\sigma||\equiv
\tr|\varrho-\sigma|$.  A similar inequality
was obtained recently by Nielsen \cite{Nielsen} for another
entanglement measure: entanglement of formation. (Nielsen
uses the Bures metric as a measure of distance.) 

Inequality (\ref{Fannes1}) shows that relative entropy of
entanglement has very regular asymptotic behaviour, and is
a suitable parameter to describe asymptotic  manipulations
of entanglement. In particular, it is easy to see that the
inequality guarantees the continuity of the form
(\ref{slabsza}). Our proof is also valid for variations of
the considered measure (for example, if, as in
\cite{Rains_bound},  we minimize over positive partial
transpose states rather than over separable states). Note
here that it is not clear whether one should expect such
strong continuity for all relevant parameters in quantum
entanglement theory. For example, distillable 
entanglement  satisfies the weaker continuity
(\ref{slabsza}) for pure
$\varrho'$, but might not satisfy the inequality
(\ref{Fannes1}) e.g. near  the border between bound
entangled and free entangled states
\cite{bound,transp}.  It should also be noted that, even on
finite dimensional Hilbert spaces, relative entropy itself
is not a continuous function so that continuity of related
functions cannot be taken for granted.  Relative entropy is
however lower semicontinuous
\cite{otre,Ohya} --- on convergent sequences of states, it
can jump down but not up.
\medskip

\noindent{\bf Theorem }{\sl Let $\dcal$  be a set of states
(density matrices)  on a Hilbert space $\hcal$ of
dimension $N < \infty$. Suppose that
$\dcal$ is a compact convex set which includes the
maximally chaotic state
$\tau\equiv {I\over N}$. Then the function given by
\[ E(\sigma) =\inf_{\varrho\in\dcal}S(\sigma|\varrho), \quad
\mbox{where}\quad S(\sigma|\varrho)=\tr
\sigma\log\sigma-\tr\sigma\log\varrho
\] satisfies the inequality
\be |E(\sigma_1)-E(\sigma_2)|\leq 2
(||\sigma_1-\sigma_2||\log N
+\eta(||\sigma_1-\sigma_2||))+4||\sigma_1-\sigma_2||
\label{main_ineq}
\ee for $||\sigma_1-\sigma_2||\leq {1\over 3}$.}
\medskip

\noindent{\bf Remarks } (i) If $\hcal=\hcal_A\otimes
\hcal_B$  and $\dcal$ is the set of separable
(disentangled) states, then  $E$ is the relative entropy of
entanglement. If, instead, $\dcal$ is the set of matrices
with positive partial transposition \cite{Peres,transp}
then we obtain  Rains
\cite{Rains_bound} bound for distillable entanglement. (ii)
Note that the inequality (\ref{main_ineq}) can be written
in the form (\ref{Fannes1}).
\medskip

\noindent {\bf Proof }
 For any state $\sigma$, let $\hat \varrho(\sigma) \in
\dcal$ denote a state such that  $E(\sigma) =
S({\sigma}|{\hat \varrho(\sigma)})$.  
$\hat \varrho(\sigma)$ exists because $\dcal$ is compact
and $S$ is lower semicontinuous. Let  $S_1(\sigma) =
-\tr(\sigma \log \sigma)$ be von Neumann entropy. The
theorem is clearly true if $\sigma_1 = \sigma_2$, so
suppose otherwise. For $0 < x \leq 1$, let $E_x(\sigma) =
\inf\{ S(\sigma|x\varrho + (1-x)\tau):
\varrho \in
\dcal\}$. Choose $\hat \varrho_x(\sigma) \in \dcal$ such
that $E_x(\sigma) = S(\sigma|x\hat
\varrho_x(\sigma) + (1-x)\tau)$. By the monotonicity of the
logarithm,
\begin{eqnarray*}
 S(\sigma|x\varrho + (1-x)\tau) \  &&= -S_1(\sigma) -
\tr(\sigma \log(x
\varrho + (1-x) \tau))\nonumber \\ &&\leq -S_1(\sigma) -
\tr(\sigma \log
\varrho) -
\log x = S(\sigma|\varrho) - \log x.
\end{eqnarray*}
 Thus
\[ E(\sigma) - \log x = S(\sigma|\hat \varrho(\sigma)) -
\log x \geq S(\sigma|x\hat \varrho(\sigma) + (1-x)\tau)
\geq E_x(\sigma) \geq E(\sigma).
\] Set  $x = 1 -  ||\sigma_1 - \sigma_2||$.  By the simple
inequality
$|\log x|\leq 2(1-x)$ for ${1\over2} \leq x \leq 1$, we have
$|\log x | \leq 2 ||\sigma_1 - \sigma_2||$ so that
$|E(\sigma) - E_x(\sigma)| \leq 2 ||\sigma_1 - \sigma_2||$.
Now we shall  use Fannes' inequality  \cite{Fannes}
 \be |S_1(\sigma_1) - S_1(\sigma_2)| \leq ||\sigma_1 -
\sigma_2|| \log N +
\eta(||\sigma_1 -
\sigma_2||)
\label{Fannes2}
\ee which holds for  $||\sigma_1-\sigma_2||\leq {1\over3}$.
The monotonicity of the logarithm gives
\[ 0 \geq \log(x \varrho + (1-x)
\tau) \geq - \log N + \log(1-x).
\] Composing this with the standard inequality \cite{Sakai}
\[ |\tr(\sigma_1 A) - \tr(\sigma_2 A)|
\leq ||\sigma_1 - \sigma_2||\ ||A||_{op}
\] (where $||\cdot||_{op}$ denotes operator norm),  which
holds for any operators $\sigma_1$, $\sigma_2$ and $A$ in
finite  dimensions, we obtain
\begin{eqnarray} |\tr(\sigma_1 &&\log(x \varrho + (1-x)
\tau)) -
\tr(\sigma_2\log(x \varrho + (1-x) \tau))|\nonumber \\ &&
\leq ||\sigma_1 - \sigma_2||\, ||\log(x \varrho + (1-x)
\tau)||_{op}
\nonumber \\ && \leq ||\sigma_1 - \sigma_2||(\log N -
\log(1-x)) = ||\sigma_1 - \sigma_2||\log N +
\eta(||\sigma_1 - \sigma_2||).
\label{trace}
\end{eqnarray}  
{}From the inequalities (\ref{Fannes2}) and (\ref{trace}) it
follows that
\[ |S(\sigma_1|x\varrho + (1-x)\tau) - S(\sigma_2|x\varrho +
(1-x)\tau)|
\leq 2(||\sigma_1 - \sigma_2|| \log N + \eta(||\sigma_1 -
\sigma_2||)).
\] Then
\begin{eqnarray*} E_x(\sigma_1)\  &&= S(\sigma_1|x\hat
\varrho_x(\sigma_1) + (1-x)\tau) \leq S(\sigma_1|x\hat
\varrho_x(\sigma_2) + (1-x)\tau)
\\ && \leq S(\sigma_2|x\hat \varrho_x(\sigma_2) +
(1-x)\tau) + 2(||\sigma_1 - \sigma_2|| \log N +
\eta(||\sigma_1 - \sigma_2||))
\\ &&= E_x(\sigma_2) + 2(||\sigma_1 - \sigma_2|| \log N +
\eta(||\sigma_1 - \sigma_2||))
\end{eqnarray*}
 and, by symmetry
\[ |E_x(\sigma_1) - E_x(\sigma_2)| \leq 2(||\sigma_1 -
\sigma_2|| \log N +
\eta(||\sigma_1 - \sigma_2||)).
\]  Finally,
\begin{eqnarray*} |E(\sigma_1) - E(\sigma_2)| \  && \leq 
|E(\sigma_1) - E_x(\sigma_1)| + |E_x(\sigma_1) -
E_x(\sigma_2)| + |E_x(\sigma_2) - E(\sigma_2)| 
\\ &&\leq  2(||\sigma_1 - \sigma_2|| \log N +
\eta(||\sigma_1 - \sigma_2||)) + 4 ||\sigma_1 - \sigma_2||. 
\end{eqnarray*}
\hbox to \textwidth{\hfill \blacksquare}
\medskip

It is often suggested that we interpret $\hat
\varrho(\sigma)$ as the state in $\dcal$ closest to
$\sigma$.  This suggestion is not entirely unproblematic. 
For example, even if $\dcal$ is the separable states, there
are $\sigma$ for which $\hat \varrho(\sigma)$ is not
unique.  As a final result, therefore, we use our theorem
and standard results about relative entropy
(\cite{otre,Ohya}) to prove a non-obvious property of
$\hat \varrho(\sigma)$, which is essential to this
interpretation.
\medskip

\noindent{\bf Corollary }{\sl Let $\sigma_n$  be a sequence
of states converging to a state $\sigma$ which is in
$\dcal$.  Then $\hat
\varrho(\sigma_n)$ also converges to $\sigma$.}
\medskip

\noindent {\bf Proof } Suppose not.  By compactness, there
is a subsequence such that
$\hat \varrho(\sigma_{n_k}) \rightarrow \varrho \ne
\sigma$.  By the theorem $E(\sigma_n) \rightarrow 0$.  But
\[ \hfill \liminf E(\sigma_{n_k}) = \liminf
S(\sigma_{n_k}|\hat
\varrho(\sigma_{n_k})) \geq S(\sigma|\varrho) > 0. \hfill
\blacksquare \]
\medskip

\noindent{\bf Acknowledgements } This work was made
possible by the Cambridge Newton Institute Programme
``Communication, Complexity and Physics of Information''
(1999), supported by the European Science Foundation. M.H.
acknowledges the Polish Committee for Scientific Research,
contract No. 2 P03B 103 16.

\end{document}